\documentclass[twocolumn,superscriptaddress,prl,nofootinbib]{revtex4-2}

\usepackage{footnote}
\usepackage[titletoc,toc,title]{appendix}
\usepackage{titletoc}
\usepackage{titlesec}
\usepackage[version=4]{mhchem} 
\usepackage[normalem]{ulem}
\usepackage{mhchem}
\usepackage{amsmath}
\usepackage{placeins}
\usepackage{amssymb}
\usepackage{amsfonts}
\usepackage{graphicx}
\usepackage{xcolor}
\usepackage{xfrac}
\usepackage{footmisc}
\usepackage{comment}
\usepackage{pifont}
\usepackage{times}
\usepackage[hidelinks]{hyperref}
\usepackage{enumitem}
\usepackage{centernot}
\usepackage{cancel}
\usepackage{slashed}
\usepackage{relsize}
\usepackage{amsmath,amssymb,mathrsfs}
\usepackage{times}
\usepackage{epsfig}
\usepackage{verbatim}
\usepackage{bm}
\usepackage[utf8]{inputenc}
\usepackage{graphics}
\usepackage{graphicx,epsfig,amssymb,amsmath,color,cancel}
\usepackage{subfigure}
\usepackage[english]{babel}
\usepackage{adjustbox}
\usepackage{physics}

\usepackage{array}
\newcolumntype{P}[1]{>{\centering\arraybackslash}p{#1}}
\newcolumntype{M}[1]{>{\centering\arraybackslash}m{#1}}

\definecolor{zima_blue}{HTML}{1393C1}
\hypersetup{setpagesize=false,bookmarksnumbered=true,bookmarksopen=true,%
colorlinks=true,linkcolor=zima_blue,urlcolor=zima_blue,citecolor=zima_blue,linktocpage=false}

\usepackage{breakcites}

\def\MPl		{M_{\mathsmaller{\rm Pl}}}

\newcommand{\M}{\mathsmaller{\rm M}}
\newcommand{\DM}{\mathsmaller{\rm DM}}

\newcommand{\RH}{\mathsmaller{\rm RH}}

\newcommand{\Tnuc}{T_{\rm n}}

\newcommand{\Teq}{T_{\rm eq}}

\newcommand{\cvac}{c_{\mathsmaller{\rm vac}}}

\def\grun		{\gamma_{\rm run}}

\def\PLO        {\mathcal{P}_{\mathsmaller{\rm LO}}}

\def\PLL        {\mathcal{P}_{\mathsmaller{\rm LL}}}

\def\Pheav      {\mathcal{P}_{{\rm heavy}}}
\def\gLL        {\gamma_{\mathsmaller{\rm LL}}}
\def\gcoll      {\gamma_{{\rm coll}}}
\def\Teq        {T_{\rm eq}}

\usepackage{tikz-feynman}

\usepackage{empheq}
\usepackage[most]{tcolorbox}
\newtcbox{\mymath}[1][]{%
    nobeforeafter, math upper, tcbox raise base,
    enhanced, colframe=blue!20!black,
    colback=blue!15, boxrule=1pt,
    #1}

\begin{document}

\title{\bf{\fontsize{17pt}{0}\selectfont  Bubbletrons}\\\vspace{0.2cm} {\fontsize{12pt}{0}\selectfont Ultrahigh-Energy Particle Collisions and Heavy Dark Matter at Phase Transitions}}

\author{Iason Baldes}
\email{iasonbaldes@gmail.com}
\affiliation{Laboratoire de Physique de l'\'Ecole Normale Sup\'erieure, ENS, \\ Universit\'e PSL, CNRS, Sorbonne Universit\'e, Universit\'e Paris Cit\'e, F-75005 Paris, France}
\author{Maximilian Dichtl}
\email{maximilian.dichtl@lpthe.jussieu.fr }
\affiliation{Laboratoire de Physique Th\'eorique et Hautes \'Energies, \\ CNRS, Sorbonne Universit\'e, F-75005 Paris, France}
\affiliation{Dipartimento di Fisica e Astronomia, Università di Bologna
and INFN sezione di Bologna, Via Irnerio 46, I-40126 Bologna, Italy}
\author{Yann Gouttenoire}
\email{yann.gouttenoire@gmail.com}
\affiliation{School of Physics and Astronomy, Tel-Aviv University, Tel-Aviv 69978, Israel}
\affiliation{PRISMA+ Cluster of Excellence $\&$ MITP, Johannes Gutenberg University, 55099 Mainz, Germany}
\author{Filippo Sala}
\email{filo.sala@gmail.com}
\thanks{FS is on leave of absence from LPTHE, CNRS \& Sorbonne Universit\'{e}, Paris, France. \\}
\affiliation{Dipartimento di Fisica e Astronomia, Università di Bologna
and INFN sezione di Bologna, Via Irnerio 46, I-40126 Bologna, Italy}

\begin{abstract}

We initiate the study of `bubbletrons', by which we mean ultra-high-energy collisions of the particle shells that generically form at the walls of relativistic bubbles in cosmological first-order phase transitions (PT).
As an application, we calculate the maximal dark matter mass $M_\DM$ that bubbletrons can produce in a $U(1)$ gauge PT, finding $M_\DM \sim 10^5/10^{11}/10^{15}$~GeV for PT scales $v_\phi \sim 10^{-2}/10^3/10^9$ GeV. Bubbletrons realise a novel link between ultra-high-energy phenomena and gravitational waves (GW) sourced at the PT, from nanohertz to megahertz frequencies.

\vspace{0.4cm}
\noindent
DOI:~\href{https://doi.org/10.1103/PhysRevLett.134.061001}{ 10.1103/PhysRevLett.134.061001}

\end{abstract}

\maketitle

\section{Introduction}
Particle accelerators of different sorts continue to play a prominent role in physics. Laboratory accelerators gave us an immense amount of knowledge about the fundamental building blocks of Nature. Astrophysical accelerators (supernovae, active galactic nuclei, ...), furthermore, contributed not only to our understanding of the universe, but also shaped the way it looks. In this letter we point out that cosmological particle accelerators may also have existed, if a first order phase transition (PT) took place in the early universe, and we begin to quantitatively explore their implications. 

Along with the electroweak and QCD transitions, known to be crossovers in the Standard Model~\cite{Kajantie:1996mn,Aoki:2006we}, one or more first order PTs may have taken place in the first second after inflation.
They are commonly predicted in motivated extensions of the SM, such as extra-dimensional~\cite{Creminelli:2001th}, confining~\cite{Nardini:2007me,Konstandin:2011dr}, or supersymmetric models~\cite{Craig:2020jfv}, and solutions to the strong CP~\cite{DelleRose:2019pgi,VonHarling:2019rgb}, flavour~\cite{Greljo:2019xan}, or neutrino mass problems~\cite{Jinno:2016knw}. 
Independently of where they come from,
such PTs may have far-reaching consequences through the cosmological relics they can leave behind, e.g.~primordial black holes~\cite{Hawking:1982ga,Kodama:1982sf,Liu:2021svg,Hashino:2021qoq,Kawana:2022lba,Lewicki:2023ioy,Gouttenoire:2023naa,Baldes:2023rqv,Gouttenoire:2023bqy,Gouttenoire:2023pxh,Gross:2021qgx,Kawana:2021tde}, topological defects~\cite{Aharonov:1959fk,Nielsen:1973cs,Kibble:1976sj,Gouttenoire:2019kij,Gouttenoire:2019rtn}, magnetic fields~\cite{Hogan:1983zz,Quashnock:1988vs,Vachaspati:1991nm,Enqvist:1993np,Sigl:1996dm,Ahonen:1997wh,Ellis:2020nnr}, dark matter (DM)~\cite{Watkins:1991zt, Chung:1998ua,Falkowski:2012fb,Hambye:2018qjv,Baldes:2018emh,Baker:2019ndr,Chway:2019kft,Baldes:2020kam,Azatov:2021ifm,Baldes:2021aph,Kierkla:2022odc,Freese:2023fcr,Gouttenoire:2023roe,Giudice:2024tcp}, the baryon asymmetry~\cite{Kuzmin:1985mm,Shaposhnikov:1986jp,Cohen:1990py,Shaposhnikov:1991cu,Farrar:1993sp,Huet:1994jb,Gavela:1994dt,Morrissey:2012db,Konstandin:2013caa,Servant:2018xcs,Katz:2016adq,Azatov:2021irb,Baldes:2021vyz,Dichtl:2023xqd}, together with a background of gravitational waves (GW)~\cite{Witten:1984rs,Kosowsky:1992rz,Kamionkowski:1993fg,Randall:2006py,Huber:2008hg,Hindmarsh:2013xza,Jinno:2017fby,Konstandin:2017sat,Cutting:2018tjt,Lewicki:2020jiv}.

As the universe expands, sitting in its lowest free energy vacuum, another vacuum may develop at a lower energy due to the fall in temperature, eventually triggering a PT. If a PT is first order then it proceeds via the nucleation of bubbles of broken phase into the early universe bath (see e.g.~\cite{Hindmarsh:2020hop,Gouttenoire:2022gwi} for reviews). Bubble walls that expand with ultrarelativistic velocities store a lot of energy, locally much higher than both the bath temperature and the scale of the PT. %
Wall interactions with the bath then necessarily accelerate particles to high energies and accumulate them into shells, as first worked out in specific cases in~\cite{Baldes:2020kam,Gouttenoire:2021kjv,Baldes:2022oev,GarciaGarcia:2022yqb}.
Collisions of shells from different bubbles constitute a ultra-high-energy collider in the early universe, which we dub `bubbletron'~\footnote{Bubbletrons are not to be confused with the idea of testing new particles (lighter than Hubble) via their imprint on primordial non-gaussianities, which was named `cosmological collider'~\cite{Arkani-Hamed:2015bza} by a possible analogy with laboratory colliders, but where actually no acceleration mechanism is in place.}.

In this letter we initiate a quantitative study of bubbletrons.
We review wall velocities in Sec.~\ref{sec:walls} and determine the shells' collision energies in Sec.~\ref{sec:energies}, we classify bubbletrons and calculate the resulting production of heavy particles in Sec.~\ref{sec:shells}, apply our findings to heavy DM and correlate them with the GW from the PT in Sec.~\ref{sec:DM}. In Sec.~\ref{sec:outlook} we conclude.

\section{First order phase transitions with relativistic bubble walls}
\label{sec:walls}

We consider a cosmological first-order PT between two vacuum states with zero-temperature energy density difference $\Delta V = c_{\rm vac}\,v_\phi^4$, where $c_{\rm vac}  \lesssim \mathcal{O}(1)$ is a model-dependent parameter and $v_\phi$ is the VEV of the PT order parameter $\phi$ (e.g. a scalar field) in the final vacuum.
As the universe expands and cools its temperature falls below the critical temperature $T_c$, i.e.~when the two minima of the thermal potential have the same free energy density, and the PT becomes energetically allowed.
The PT happens around the temperature $T=T_n  < T_c$, defined by the condition $\Gamma(T_n) = H^4(T_n)$, where $H$ is the Hubble parameter and $\Gamma$ the tunneling rate, per unit volume, between the two vacua. 
At $T_n$ bubbles of the broken phase (i.e.~where the order parameter sits in its zero-temperature vacuum) are nucleated and start expanding to eventually fill the universe. The time they take to do so and complete the PT is set by $\beta^{-1}$, with $\beta \equiv (d\Gamma/dt)/\Gamma$, which is shorter than a Hubble time $H^{-1}$.
%

The bubble walls are defined as the spherically symmetric regions of space where the background field $\phi$ rapidly varies, from the high-temperature value outside the bubble, to $v_\phi$ inside it. The pressure density inside is larger than outside, so the bubbles expand. If friction pressure on the walls is negligible, then they run away with a Lorentz boost
$\gamma(R) = R/(3R_\text{nuc})$~\cite{Gouttenoire:2021kjv},
where $R$ is the bubble radius and $R_\text{nuc} \approx T_n^{-1} $ is its radius at nucleation, see e.g.~\cite{Baldes:2020kam}. If the walls runaway until colliding, they reach 
\begin{equation}
\label{eq:grun}
   \grun
   = \frac{R_{\rm coll}}{3\,R_\text{nuc}}
   \simeq 2.7\cdot10^{14}
   \cdot \frac{\Tnuc}{\Teq}\, \frac{\rm TeV}{v_\phi}\, \frac{1}{(\cvac g_b)^{\!\frac{1}{4}}} \frac{20}{\beta/H} ,
\end{equation}
where $R_{\rm coll}\simeq(\pi)^{\!\frac{1}{3}}\beta^{-1}$ is their radius at collision~\cite{Enqvist:1991xw}, $\Teq = (30c_{\rm vac}/(\pi^2 g_b))^{\!\frac{1}{4}}\, v_\phi$ is the temperature when the radiation energy density, $\rho_{\rm rad} = g_b \pi^2 T^4/30$ with $g_b$ the number of relativistic degrees of freedom before the PT, equals $\Delta V$, and we have assumed $T_n \leq T_{\rm eq}$ so that $H \simeq \sqrt{\cvac/3} v_\phi^2/\MPl$.
%

A number of effects can exert pressure on walls and slow them down. 
Collisional plasma effects are expected to exert a negligible pressure for $T_n \lesssim T_{\rm eq}$ (see e.g.~\cite{Konstandin:2010dm,Cline:2021iff,Laurent:2022jrs,DeCurtis:2023hil}), which is the case we will be interested in.
One then enters the so-called ballistic regime,  where particle interactions can be neglected.
Then, one has pressure from single particles getting a mass across the wall, $\PLO = g_*\Delta m^2\,T_n^2/24$~\cite{Bodeker:2009qy}, with $g_*$ the number of degrees of freedom getting an average mass squared $\Delta m^2 \propto v_\phi^2$ at the PT. 
Another pressure that could be relevant in some models, $\Pheav$, is that from degrees of freedom heavier than $v_\phi$ that couple to the particles that feel the PT~\cite{Azatov:2020ufh}.
$\PLO$ and $\Pheav$ are both smaller than $\Delta V$ for $T_n^2 < v_\phi^2$, up to order-one model-dependent coefficients. In this case, which will be the focus of this letter, the velocity of bubble walls becomes ultrarelativistic.

Ultrarelativistic bubble walls can either run away until they collide with those of other bubbles, or reach a terminal velocity beforehand, set by yet another source of pressure given by the bremsstrahlung radiation, off bath particles, of particles that get a mass $m$ at wall crossing. If this radiation is soft-enhanced, as for emitted gauge bosons, then their pressure grows with $\gamma$~\cite{Bodeker:2017cim}. Its size is enhanced by large logarithms, that have been resummed in~\cite{Gouttenoire:2021kjv}, which gives the pressure $\PLL \simeq \frac{\zeta(3)}{\pi^4} g^2 g_{\rm eff}\,\gamma \, m_V\, T_n^3 \log(m_V/\mu)$, where $g$ is the gauge coupling, $g_{\rm eff}$ a weighted sum of the radiating degrees of freedom times their charges, $m_V$ is the gauge boson mass and $\mu$ a physical IR cut-off~\footnote{
The only two other sources of pressure, which we are aware of, are those from string fragmentation in confining PTs~\cite{Baldes:2020kam} and that from vectors that get from the wall only a small component of their mass~\cite{GarciaGarcia:2022yqb}. Neither of them applies to the scenario considered in this letter.}.
If $\PLL(\gamma)$ reaches $\Delta V = \cvac v_\phi^4$ before collision, then walls attain a terminal velocity 
\begin{equation}
\label{eq:gTV}
    \gLL \simeq
    3.5\cdot 10^4\cdot
    \Big(\frac{\Teq}{T_n}\Big)^{\!3}
    \Big(\frac{0.1}{g}\Big)^{\!\!3}
    \Big(\frac{\cvac g_b^3}{10^4}\Big)^{\!\!\frac{1}{4}}
    \frac{10}{g_{\rm eff} \log\!\frac{m_V}{\mu}},
\end{equation}
where we have chosen $m_V = g v_\phi$ for definiteness.
The typical boost of bubble walls at collision then is
\begin{equation}
\label{eq:gcoll}
{\gcoll \simeq \frac{ \gLL \grun }{ \gLL + \grun} \simeq \textrm{Min}\big[\gLL,\, \grun\big].}
\end{equation}
Large boosts at collision are realised for small gauge coupling $g$, or for large $v_\phi/T_n$, or in global (rather than gauged) PTs because there $\PLL$ does not grow with $\gamma$.

\section{Shells of particles at the walls}
\label{sec:energies}

\begin{figure}[t]
\centering
\includegraphics[width=0.35\textwidth, scale=1]{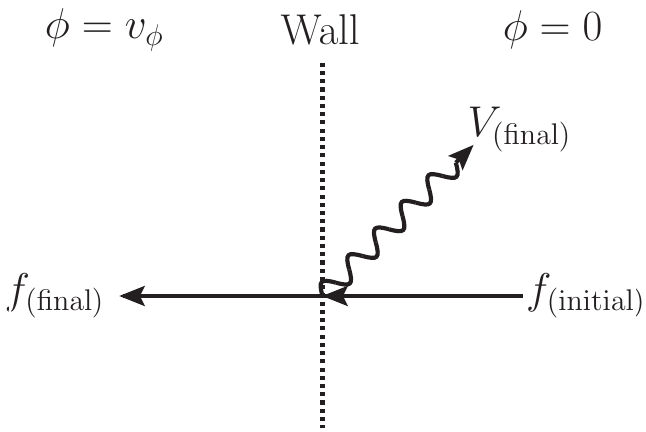}
\caption{\small The shell source considered in this letter. Charged particles entering the bubble can radiate a gauge boson which is reflected by the wall back into the false vacuum phase. These gauge bosons form shells propagating in front of the bubble walls.}
\label{fig:picture_radiated}
\end{figure}

The mechanisms at the origin of the pressures above also cause particles to accumulate into shells, which we list below:
\begin{enumerate}
\item
Particles acquiring their mass \cite{Bodeker:2009qy,Baldes:2021vyz};
\item
Particles radiated and transmitted in the wall \cite{Gouttenoire:2021kjv};
\item
Heavier particles if produced by lighter ones that feel the PT \cite{Azatov:2020ufh,Baldes:2021vyz,Baldes:2022oev};
\item
In confining PTs, hadrons from string fragmentation~\cite{Baldes:2020kam};
\item
Vectors acquiring a small part of their mass \cite{GarciaGarcia:2022yqb};
\item 
Particles produced by oscillations of the wall $\phi$ \cite{WallDecay}; 
\item
In confining PTs, ejecta from string fragmentation~\cite{Baldes:2020kam};
\item
Particles radiated and reflected by the wall \cite{Gouttenoire:2021kjv}.
\end{enumerate}
Shells 1 to 6 follow the bubble walls, shells 7 and 8 precede them. 
When bubbles collide, also these shells do. If their constituent particles still have a center-of-mass energy much larger than $v_\phi$ by that time, then they realise what we define a `bubbletron.' Whether that happens depends on a number of propagation effects, their study can be model-dependent and pretty complicated, and we are aware of very few attempts at carrying it out in some detail~\cite{Baldes:2020kam,Gouttenoire:2021kjv,Baldes:2022oev}. Accordingly, we have made a novel systematic study of shell propagation, that we present in another paper~\cite{Baldes:2024wuz},  because its interest goes beyond bubbletrons (for example it could affect GW from PTs), and whose results are used in this letter.


\begin{figure}[t]
\centering
\includegraphics[width=0.5\textwidth, scale=1]{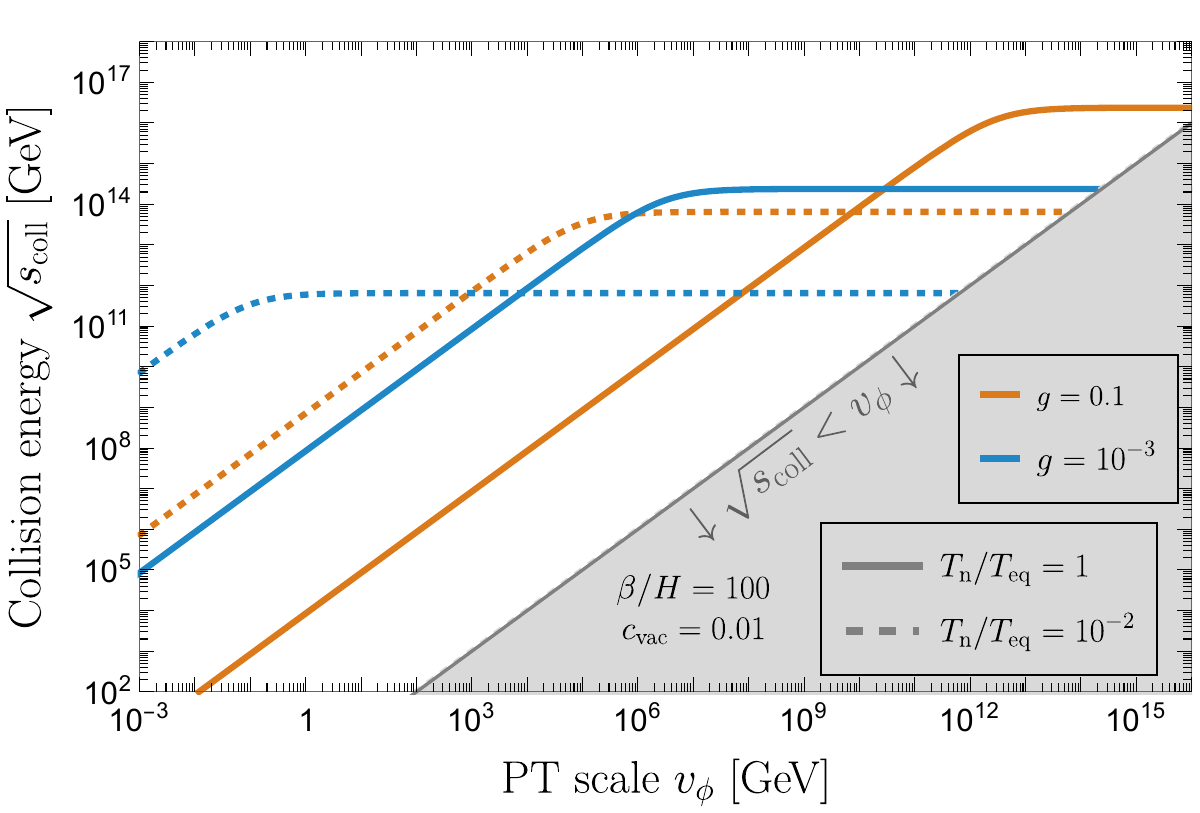}
\caption{\small Center-of-mass scattering energies of two gauge bosons radiated and reflected at the walls of different bubbles. }
\label{fig:sqrtS_vs_vphi}
\end{figure}

To give a quantitative idea of the center-of-mass energies achievable, let us consider as an example in this letter the case of radiated reflected gauge bosons \cite{Gouttenoire:2021kjv,Baldes:2024wuz} with mass $m_V$ in the broken phase. Those gauge bosons are radiated by charged particles entering bubble walls and are reflected back due to their low energy in the wall frame, see Fig.~\ref{fig:picture_radiated}. They form shells propagating in front of the bubble walls.
If those shells free stream until they collide, one has typical center-of mass collision energy squared (see Fig.~\ref{fig:sqrtS_vs_vphi})
\begin{equation}
    s_{\rm coll}
    \simeq 4\,\gcoll^2\,E_V^2
    \simeq 0.4\,\gcoll^2\,m_V^2\,,
\end{equation}
where $E_V$ is the typical energy of a reflected shell particle in the wall frame and we have assumed heads-on collisions for simplicity. In the second equality we have used $E_V^2 \simeq 0.1 m_V^2$, which we computed from the distribution $dP_V \propto \frac{dk_\perp^2}{k_\perp^2} \frac{dE_V}{E_V}(\frac{m_V^2}{m_V^2+k_\perp^2})^2 $~\cite{Gouttenoire:2021kjv}, with $k_\perp$ the component of $V$ momentum parallel to the wall.
Interestingly, collision energies can lie above the scales of both grand unification~\cite{Croon:2019kpe} and inflation~\cite{Planck:2018jri},
\begin{equation}
\sqrt{s_{\rm coll}}
\simeq 1.7\cdot10^{17}~{\rm GeV} \,
    g \,
    \frac{\gamma_{\rm coll}}{\gamma_{\rm run}}  \frac{T_{\rm n}}{T_{\rm eq}} \frac{1}{(\cvac g_b)^{\!\frac{1}{4}}} \frac{20}{\beta/H}\,.
    \label{eq:scoll_numeric}
\end{equation}


 For simplicity, from now one we consider a gauged $U(1)$ with coupling $g$ spontaneously broken by a scalar $\phi$ with charge 1. In \cite{Baldes:2024wuz}, we show that the condition that shells free stream until collision is realised for small $g$, large $v_\phi$, or large $v_\phi/T_n$. Then, for those parameters, one does obtain collisions with the high energies of Eq.~(\ref{eq:scoll_numeric}), opening up the possibility to test them with cosmology.

\section{Shell collision products}
\label{sec:shells}

We assume that a collision of particles $i$ and $j$ from two different shells produces one much heavier particle $Y$ with cross section times Moeller velocity $\sigma v_\M|_{ij}(s)$.
The probability that a particle $i$ undergoes one such interaction is given by $\int_{\mathcal{\lambda}=0}^{L_{\rm shell}}\frac{d\lambda}{v_w} \,n_j(\lambda)\, \sigma v_\M|_{ij}(s(\vec{x}_i,\lambda))$, where $L_{\rm shell}$ is the length of the shell of particles $j$, $v_w$ the speed of the wall, $\lambda$ a radial coordinate and $\vec{x}$ a space one, and $s$ depends on them because particles in different layers in a shell have different energy (e.g. in the run-away regime particles reflected later are more energetic, because $\gamma$ grows with $R$).
Spherical symmetry of the layers implies $d\lambda = d^3\vec{x}_j/(4\pi R_{\rm coll}^2)$. We can then multiply by the total number of particles $i$, $N_i = \int_{\rm shell1} \! d^3\vec{x}_i\,n_i(\vec{x})$, and using $v_w \simeq 1$ and $v_\M \simeq 2$,  write the total number of $Y$ produced as
\begin{equation}
\label{eq:NYcoll}
N_Y
= \frac{N_{\rm shells}}{4\pi R_{\rm coll}^2} \int_{\rm shell\,1} \!\!\!\!\!\!\!\!\! d^3\!\vec{x}_i \,n_i(\vec{x}_i) \!
\int_{\rm shell\,2} \!\!\!\!\!\!\!\!\! d^3\!\vec{x}_j \,n_j(\vec{x}_j)\, \sigma_{ij}(s(\vec{x}_i,\vec{x}_j)),
\end{equation}
where $N_{\rm shells}$ is the total number of shells (i.e.~of bubbles) that collide and we have divided by 2 to avoid double-counting the initial $i,j$ particles when summing over all shells.
Let us now write for simplicity $s(\vec{x}_i,\vec{x}_j) \simeq s_{\rm coll}$, which is an excellent approximation in the terminal velocity regime, and only overestimates $s$ by an $\mathcal{O}(1)$ factor in the run-away one. Then we can take $\sigma_{ij}^{\rm coll} \equiv \sigma_{ij}(s_{\rm coll})$ out of the integrals, and write the average number density of $Y$ from collisions as
\begin{equation}
n_Y \equiv \frac{N_Y}{V_{\rm uni}}
\simeq  N_{\rm shells}\frac{N_i N_j \sigma_{ij}^{\rm coll}}{V_{\rm uni} 4\pi R_{\rm coll}^2}
=\frac{N_{\rm b}^2 P_{b\to i} P_{b \to j} \sigma_{ij}^{\rm coll}}{N_{\rm shells} V_{\rm uni} 4\pi R_{\rm coll}^2}\,,
\end{equation}
where $V_{\rm uni}$ is the spatial volume of the universe, $N_b = n_b V_{\rm uni}$ is the number of bath particles in the entire universe and $P_{b\to i,j}$ the probability that they produce one particle $i$ or $j$ upon encountering a bubble (which is independent of $x$). 
In the last equality we have used $N_{i,j}=N_b P_{b\to i,j} V_{\rm bubble}/V_{\rm uni}=N_b P_{b\to i,j}/N_{\rm shells}$. We now use $V_{\rm bubble} = 4\pi R_{\rm coll}^3/3$ to finally write the Yield
\begin{equation}
\label{eq:Y_general}
Y_Y
\equiv \frac{n_Y}{s_\RH}
= \frac{1}{s_\RH} n_b^2 P_{b \to i} P_{b \to j} \sigma_{ij}^{\rm coll}\, \frac{R_{\rm coll}}{3}\,,
\end{equation}
where $s_\RH$ is the entropy density at reheating after the PT.
This result is valid as long as $Y\bar{Y} \to ij$ is not efficient, we checked this holds in the parameter space of our interest.


The discussion above applies to any bubbletron, including those where different populations are colliding. For concreteness, we now specify it to the case of a gauged $U(1)$, with $i=j = V$, for which~\cite{Gouttenoire:2021kjv} 
\begin{equation}
    P_{b \to V}
    \simeq \frac{g_{\rm emit}}{g_b}\frac{g^2}{16\pi^2} \log_V^2, \quad
    \log_V^2 = \log{\frac{m_V^2}{\mu^2}} 
    \Big(\log{\frac{m_V^2}{\mu^2}} - 2\Big),
\end{equation}
 where $g_b$ is the number of relativistic degrees of freedom in the bath and $g_{\rm emit}$ is the subset charged under $U(1)$, which can thus emit a vector boson $V$. For reference, in our figures we use $g_{\rm emit} = g_{\rm eff} = 10$ and $g_b = 106.75+g_{\rm emit}$. Here $\mu$ is an IR cut-off which, dealing with an abelian theory, we take as the thermal mass $\mu^2 \simeq g_{\rm emit} g^2  T_{n}^2/10$. In principle one should also include the screening length due to the high density of particles in the shell (see e.g.~\cite{Baldes:2020kam}), but the $V$'s are $U(1)$ singlets and so do not contribute at this order, and the density of fermions or scalars in the shell is suppressed, with respect to $n_V$, by extra powers of $g^2$ or $1/\gcoll$.
 We assume further that a heavier fermion $Y$ with charge $q_Y$ under the $U(1)$ exists in the spectrum.
 We compute the $Y\bar{Y}$ production cross section as
\begin{equation}
    \sigma_{VV \to Y\bar{Y}} 
    = \dfrac{q_Y^4 g^4}{4 \pi s} \, f_{Y\bar{Y}}
    \xrightarrow[s \gg M_Y^2]{} \frac{q_Y^4 g^4}{4 \pi s} \Big(\log\frac{s}{M_Y^2}-1\Big)\,,
\end{equation}
where in figures and numerical results we use the full expression $f_{Y\bar{Y}}(y\equiv \frac{4 m_{Y}^2}{s}) = ( -\sqrt{1-y}(1+y) + ( 2 + (2-y)y ) \tanh^{-1}(\sqrt{1-y}))$.
Using Eq.~\eqref{eq:gcoll}, 
$m_V = g v_\phi$, $n_b = g_b \zeta(3) T_n^3/\pi^2$ and $s_\RH = g_\RH 2 \pi^2 T^3_\RH/45$, with $T_\RH ={(1+T_n^4/T_{\rm eq}^4)^\frac{1}{4}}(30 \cvac/(g_\RH\pi^2))^\frac{1}{4} v_\phi$ the reheating temperature and $g_\RH$ the number of relativistic degrees of freedom after the PT, we find {for $s_{\rm coll} \gg  4 m_{Y}^2$}

\begin{figure}[t]
\centering
\includegraphics[width=0.5\textwidth, scale=1]{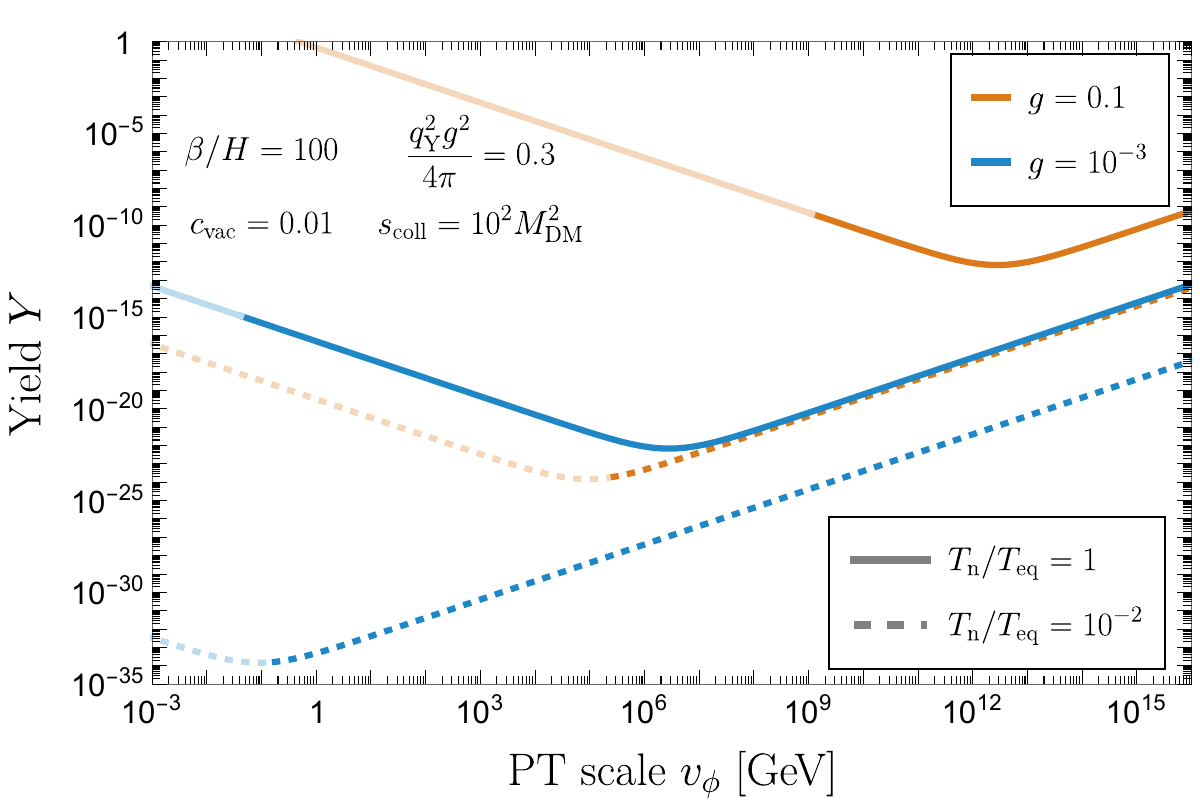}
\caption{Yield of secondary particles $Y$ produced from collision of radiated reflected $U(1)$ gauge bosons $V$. Lines become lighter when {the shell} free streaming conditions are not respected~\cite{Baldes:2024wuz}, {the shell properties change before collision}, and thus our derivation of the yield should be changed.
}
\label{fig:Y_vs_vphi}
\end{figure}

\begin{widetext}
\begin{equation}
\label{eq:Y_U1}
Y_{Y+\bar{Y}}^{U(1)}
\simeq
2.0 \cdot 10^{-20} g^2
{\left(1+\frac{T_n^4}{T_{\rm eq}^4}\right)^{\!-\frac{3}{4}}}\!\!\Big(\frac{T_n}{\Teq}\Big)^{\!4}
\Big(\frac{\grun}{\gcoll}\Big)^{\!2} \frac{v_\phi}{\rm TeV} \frac{\beta/H}{20}
\frac{g_{\rm emit}^2}{g_b}
\Big(\frac{q_Y^2 g^2/4\pi}{0.1}\Big)^2
\Big(\frac{\cvac}{0.1}\Big)^{\!\frac{3}{4}}
\Big(\frac{100}{g_\RH}\Big)^{\!\frac{1}{4}}
\frac{f_{Y\bar{Y}} \log_V^4}{100}
\,.
\end{equation}
\end{widetext}
$Y_{Y+\bar{Y}}^{U(1)}$ is visualized in Fig.~\ref{fig:Y_vs_vphi} for some representative values of the parameters.
We stress that Eqs.~(\ref{eq:Y_U1}), and the more general one~(\ref{eq:Y_general}), apply only in regions of parameter space where the free-streaming conditions of~\cite{Baldes:2024wuz} are satisfied. 
We display this in Fig.~\ref{fig:Y_vs_vphi} by interrupting the lines of $Y_Y^{U(1)}$ as soon as the free-streaming conditions are violated. 
Our calculations have potentially wide applications, which we begin to explore here for the production of heavy dark matter.

\begin{figure}[b]
\centering
\includegraphics[width=0.5\textwidth, scale=1]{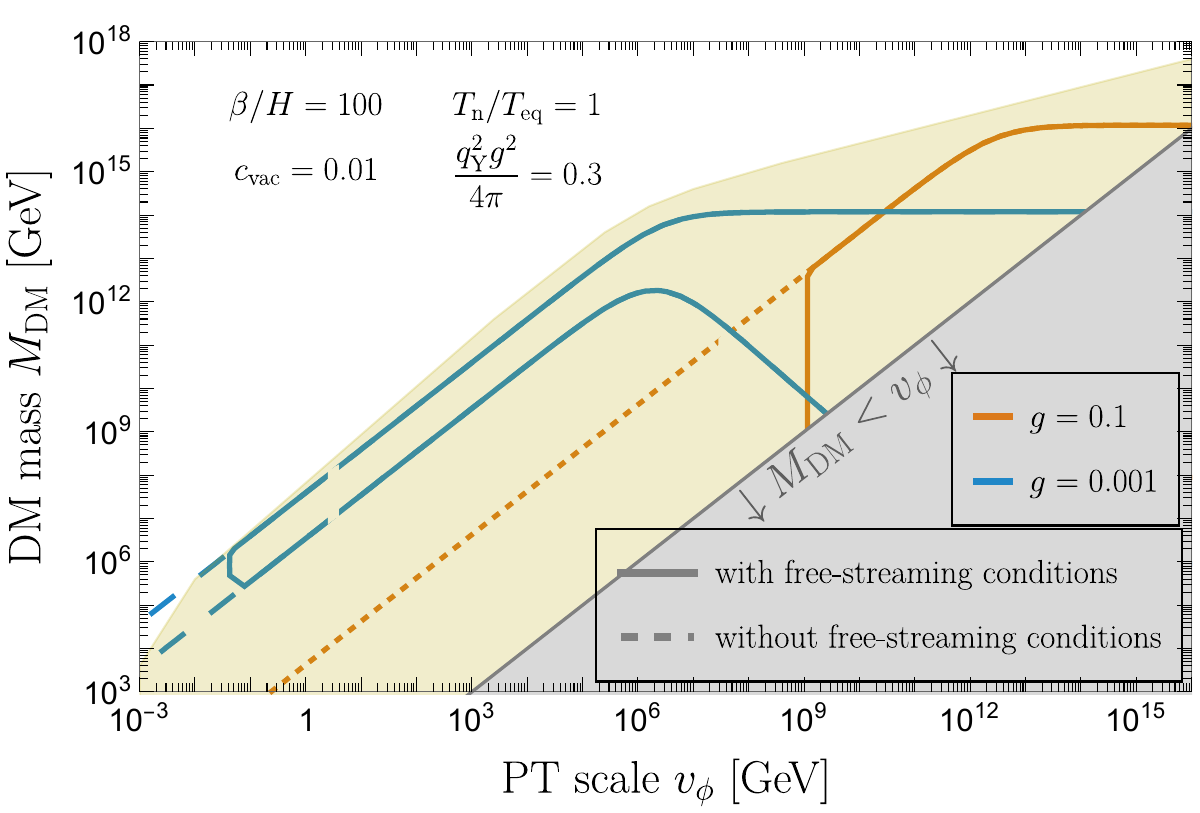}
\caption{
Mass of dark matter produced by a bubbletron of reflected $U(1)$ gauge bosons.
Lines: $M_\DM$ in two representative scenarios, both with $\beta/H = 100$ and $T_n=\Teq$. Lines turn dashed when they stop satisfying the condition of shells free streaming \cite{Baldes:2024wuz}. The champagne shaded area is the envelope of all the solid lines we obtain upon varying $g$, $T_n/\Teq$, $\beta/H$, $\cvac$, $q_Y$.
See text for more details.
}
\label{fig:max_MY_vphi_gaugeboson}
\end{figure}

\section{Heavy dark matter and gravitational waves}
\label{sec:DM}

We now specify our discussion to the case where $Y$ is stable on cosmological scales, and thus a potential DM candidate.
We assume zero initial abundance of $Y,\bar{Y}$ and impose that their yield from shell collisions reproduces the observed DM one, i.e.~$Y_{Y+\bar{Y}}^{U(1)} = Y_{\rm Planck}^\DM \simeq 0.43~{\rm eV}/M_\DM$~\cite{Planck:2018vyg} with $M_Y=M_\DM$. This allows us to plot lines of DM abundance on an $M_\DM-v_\phi$ plane, for any value of the other parameters $g$, $T_n$, etc.
    We do so varying the parameters as $1 \geq T_n/\Teq \geq 10^{-4}$, $1 \geq g \geq 10^{-5}$, $10^4 \geq \beta/H \geq 10$, $1 \geq \cvac \geq 10^{-3}$, $10^{-4} < g^2 q_Y^2/4\pi < 0.3$, with the perturbativity condition $P_{b \to V} < 1$.
We then discard all lines of DM abundance that do not satisfy the free streaming conditions of~\cite{Baldes:2024wuz}{, except the one from $VV\to\phi\phi$ in the shells plus $\phi$ momentum losses, because it does not significantly affect the momentum nor the abundance of the $V$'s (moreover one could choose $m_\phi^2 > m_V^2 + T_n^2$ to prevent it).}
The envelope of the remaining lines is represented by the champagne shaded area in Fig.~\ref{fig:max_MY_vphi_gaugeboson}. The upper edge of it gives the maximal DM mass as a function of $v_\phi$. 
For easiness of the reader, we also visualize the lines corresponding to two benchmark values of the parameters.
 One sees that in general there are two solutions that reproduce $Y_{\rm Planck}^\DM$, one for $M^2_\DM \to s_{\rm coll}/4$ and one for smaller $M_\DM$. At large $g$ the latter line falls in the region $M_\DM < v_\phi$.
At large $v_\phi$, $\gcoll = \grun$, which decreases because bubbles have less volume to expand and $M_\DM$ saturates to a constant. 
We stop the plots at $v_\phi = 10^{16}$~GeV in order to avoid the `ping-pong' regime (see e.g.~\cite{Baldes:2020kam}) where gauge bosons are reflected multiple times. At small values of $v_\phi$ the free-streaming conditions derived in \cite{Baldes:2024wuz} impose small values of $g$.

\begin{figure}[t]
\centering
\includegraphics[width=0.5\textwidth, scale=1]{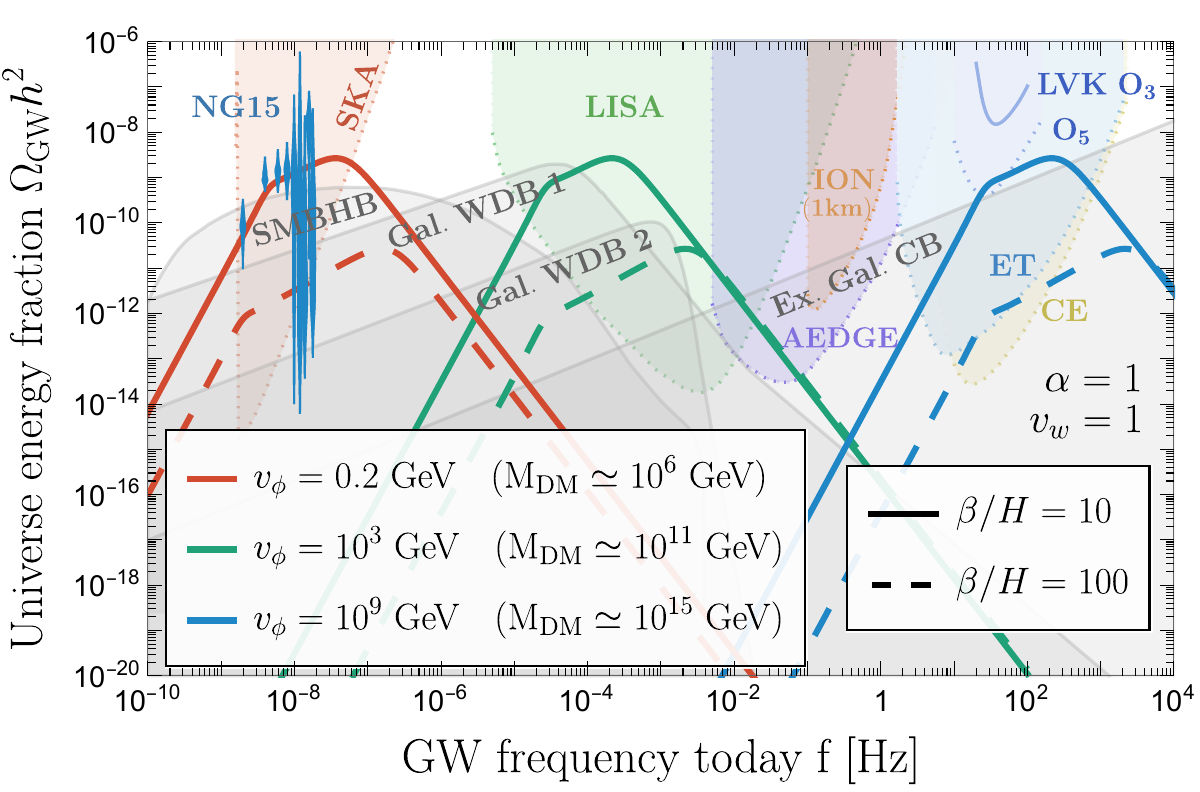}
\\[-1em]
\caption{Lines: GW signal from a $U(1)$ gauge PT. The associated bubbletron can produce dark matter with mass $M_\DM$ up to the values written in the figure for each $v_\phi$, for $\beta/H = 100$ and {$\alpha \equiv (\frac{\Teq}{T_n})^4 = 1$}.
Shaded in gray: expected foregrounds~\cite{Lamberts:2019nyk,Boileau:2021gbr,Boileau:2022ter,Robson:2018ifk,KAGRA:2021kbb}. Shaded in color: current~\cite{KAGRA:2021kbb} and projected limits~\cite{Audley:2017drz,Hild:2010id,Punturo:2010zz,Badurina:2019hst,Proceedings:2023mkp,AEDGE:2019nxb,KAGRA:2013rdx}. Blue violins:  GW signal recently detected by Pulsar Timing Arrays \cite{NANOGrav:2023gor,EPTA:2023fyk,Reardon:2023gzh,Xu:2023wog}.
}
\label{fig:GWplot}
\end{figure}

We finally compute the GW spectrum, $\Omega_{\rm GW} h^2$, generated by the PT for ultra-relativistic bubble walls using the bulk flow model~\cite{Jinno:2017fby} from~\cite{Konstandin:2017sat} for both terminal-velocity and runaway walls (see~\cite{Gouttenoire:2023bqy} for a justification), to which we impose a scaling in frequency $f^3$  for $f\lesssim H/2\pi$ as required by causality \cite{Durrer:2003ja,Caprini:2009fx,Cai:2019cdl,Hook:2020phx}. In Fig.~\ref{fig:GWplot} we display $\Omega_{\rm GW} h^2$ for three different benchmark points, where we have introduced the latent heat fraction $\alpha \equiv \Delta V/\rho_{\rm rad}|_{T_n}
= \left(T_{\rm eq}/T_{n} \right)^4$.

\section{Discussion and Outlook}
\label{sec:outlook}
In this letter we have pointed out the existence of `bubbletrons', i.e.~particle accelerators and colliders in the early universe that are generically realised by first-order phase transitions with ultrarelativistic bubble walls.
Among many processes that lead to bubbletrons, we have focused on radiated reflected particles at the walls in gauge PTs and computed their scattering energies (see Fig.~\ref{fig:sqrtS_vs_vphi}) assuming they free-stream until collision, see~\cite{Baldes:2024wuz}.
These collisions can produce particles much heavier than the scale $v_\phi$ of the PT and of inflation, with sizeable yields displayed in Fig.~\ref{fig:Y_vs_vphi}. We stress that bubbletrons are predicted in any PT with relativistic bubble walls, so they do not necessarily require vacuum domination (i.e.~$\alpha>1$ or $T_n < T_{\rm eq}$).
As an application, we found that they can produce DM
as heavy as the PeV (grand unification) scale for $v_\phi \gtrsim 10^{-2} (10^8)$~GeV, see Fig.~\ref{fig:max_MY_vphi_gaugeboson}. {Production of heavy DM can also arise from bubble-wall collisions \cite{Watkins:1991zt, Chung:1998ua,Falkowski:2012fb,Konstandin:2013caa,Freese:2023fcr,Giudice:2024tcp}. However, for first-order phase transitions that are not extremely supercooled and involve gauged sectors, only a minuscule fraction $\gLL/\grun \simeq 6\times 10^{-13}\,\alpha\times(g_b/g_{\rm eff})\cvac^{\!1/2}(v_\phi/\rm TeV) (\beta/H/100 )/g^3/ \log(m_V/\mu)$ of the latent heat $\alpha$ goes into the kinetic energy of the wall \cite{Baldes:2024wuz}. This suppression reduces the abundance of heavy DM from bubble collisions by the same factor, making the DM production from shells scattering presented here the dominant production channel.}

Our study realises a new connection between primordial GW signals and physics at energy scales otherwise inaccessible not only in the laboratory but, so far, also in the early universe. In the example of heavy DM, these GW could be accompanied by high energy cosmic rays from decays of DM, if unstable: this could intriguingly link, e.g.,
GW at pulsar timing arrays with high-energy neutrinos and photons at IceCube, KM3NeT, CTA or LHAASO. 

Our study opens several avenues of exploration.
These include bubbletrons other than those induced by radiated reflected particles, or in the region where shells do not free-stream, and applications for baryogenesis and possible trans-Planckian scatterings in the early universe. We plan to return to some of these aspects in future work.

\smallskip

\begin{acknowledgments}

{\bf Acknowledgements.}---%
YG is grateful to the Azrieli Foundation for the award of an Azrieli Fellowship.
This work was supported in part by the European Union’s Horizon 2020 research and innovation programme under grant agreement No 101002846, ERC CoG ``CosmoChart", by the Italian INFN program
on Theoretical Astroparticle Physics (TAsP), and by the French CNRS grant IEA ``DaCo: Dark Connections''.
\end{acknowledgments}

\bibliography{biblio}


\fontsize{11}{13}\selectfont




\small
\FloatBarrier

\end{document}